\documentclass[useAMS,usenatbib]{mn2e}

\usepackage{mathtext,amssymb,amsmath}
\usepackage{epsfig}
\usepackage{graphics}
\usepackage[T1]{fontenc}
\usepackage{aecompl}
\renewcommand{\vector}[1]{\ensuremath{\mathbf{#1}}}

\newcommand{\mdot}{\ensuremath{\dot{m}}}
\newcommand{\Msun}{\ensuremath{{\rm M_\odot}}}

\newcommand{\ergl}{\ensuremath{\rm erg\, s^{-1}}}

\newcommand{\keV}{\ensuremath{\rm keV}}

\newcommand{\acos}{\ensuremath{{\rm acos} \,}}

\newcommand{\pardir}[2]{\ensuremath{\frac{\partial #2}{\partial #1} }}

\newcommand{\acosh}{\ensuremath{{\rm acosh} \,}}
\newcommand{\sign}{\ensuremath{{\rm sign} \,}}

\renewcommand{\div}{\ensuremath{-}}

\title[Relativistic disc Eddington limit]{On the Eddington limit for
  relativistic accretion discs}
\author[Abolmasov \& Chashkina]{Pavel Abolmasov\thanks{E-mail:
    pavel.abolmasov@gmail.com}$^{1,2}$ and Anna Chashkina$^{1,2}$, \\
$^1$ Tuorla Observatory, Department of Physics and Astronomy, 
University of Turku, V\"ais\"al\"antie 20, FI-21500
Piikki\"o, Finland,\\
$^2$ Sternberg Astronomical Institute, Moscow State University,
 119992 Moscow, Russia\\
}
\begin{document}

\date{Accepted ---. Received ---; in
  original form --- }

\pagerange{\pageref{firstpage}--\pageref{lastpage}} \pubyear{2012}

\maketitle

\label{firstpage}

\begin{abstract}
Standard accretion disc model relies upon several assumptions, the most
important of which is geometrical thinness. Whenever this condition is
violated, new physical effects become important such as radial energy
advection and mass loss from the disc. These effects are important, for
instance, for large mass accretion rates when the disc approaches its local
Eddington limit. 
In this work, we study the upper limits for standard accretion disc
approximation and find the corrections to the standard model that should
be considered in any model aiming on reproducing the transition to
super-Eddington accretion regime. 
First, we find that for thin accretion disc, 
taking into account relativistic corrections allows to
increase the local Eddington limit by about a factor of two due to stronger
gravity in General Relativity (GR). 
 However, violation of the local Eddington limit also means large disc
  thickness. 
  To consider consequently the disc thickness effects, one should make
  assumptions upon the two-dimensional rotation law of the disc. For rotation
  frequency constant on cylinders $r\sin\theta=const$, vertical
  gravity becomes stronger with height on spheres of constant radius. On the
  other hand, effects of radial flux advection increase the flux density in
  the inner parts of the disc and lower the Eddington limit. In general, the
  effects connected to disc thickness tend to increase the local Eddington
  limit even more. The efficiency of accretion is however decreased by
  advection effects by about a factor of several. 
\end{abstract}


\section{Introduction}

Standard disc accretion introduced by~\citet{SS73} successfully explains the
thermal component of the spectra of X-ray binaries \citep{done07}. 
It is also applied to
explain the ``blue bumps'' of the spectral of active galactic nuclei but here
the success of the model is less certain (see for instance~\citet{lawrence}). Standard disc model
relies on several assumptions, the most important of them connected to the
``thinness'' of the disc meaning that the motion of the accreted matter does
not differ significantly from Keplerian rotation in a plane passing through
the accretor. Equations of vertical and radial structure may be decoupled then
due to large differences in spatial scales and gradients in these two distinct
directions. 
Thinness also implies ``locality'' condition meaning that all the
energy dissipated at some radius is radiated away at the same radius (because
vertical radiation transfer is much faster). 

The basic assumptions of the thin disc model are violated at low and high mass
accretion rates. First is connected with the cooling time scales that become
very
large for low densities \citep{evaporation}. At large mass accretion rates, on one
hand, the cooling timescales again become high because of the very high
optical depths, while on the other hand, the radiation flux generated inside
the disc becomes strong enough for the accretion disc to approach its
Eddington limit. 

When \citet{eddlimit} introduced the luminosity limit, he adopted
spherical symmetry. But the geometry of a real astrophysical system may be
different. The physics underlying the Eddington luminosity is force
balance between gravity and radiation pressure determined primarily by the
radiation flux\footnote{There are uncertainties connected to limb darkening and
  radiation collimation that we do not consider in this paper.}. 
Above a certain photosphere
belonging to a star or an accretion disc, the limiting value for radiation
flux is:
\begin{equation}
F_{\rm Edd} = \dfrac{c}{\varkappa} g,
\end{equation}
where $\varkappa$ is opacity (usually set to Thomson electron scattering
opacity that equals $\varkappa\simeq 0.34{\rm cm^2\,g^{-1}}$ for Solar metallicity), 
$c$ is the speed of light and $g$
is the gravity in the direction normal to the photosphere. Spherical symmetry
in classical mechanics implies $F = L/4\pi R^2$ and $g=GM/R^2$ and hence
there is a luminosity limit independent of distance:
\begin{equation}\label{E:eddlimit}
L_{\rm Edd} = \dfrac{4\pi GMc}{\varkappa} \simeq 1.4\times 10^{38}
\dfrac{M}{\Msun} \ergl .
\end{equation}
If the luminosity is released by accretion processes, the upper limit for
luminosity implies an upper limit for the mass accretion rate (as radiation
pressure opposes matter infall) and hence:
\begin{equation}
\dot{M}_{\rm max}\simeq \dfrac{1}{\eta}
\dfrac{L_{\rm Edd}}{c^2} ,
\end{equation}
where $\eta$ is accretion efficiency (radiation energy released per
unit accreting mass). Again, such a limit assumes spherical symmetry. 

On the other hand, in accretion discs, gravity (compensated by centrifugal
force in radial direction) and radiation, both directed roughly vertically,
depend on both the vertical and radial coordinates in the disc
hence the radiation pressure limit should be applied locally. 
Vertical gravity becomes stronger with height, hence increase in local energy
release may be compensated by increased disc thickness (we will hereafter use
half-thickness $H$) up to a certain thickness comparable to the local radial
coordinate ($H\sim R$) where the thin-disc
approximation breaks down. Above this accretion rate value that we will call
``critical'', accretion disc structure is significantly distorted inside the
outermost radius where the disc becomes thick (spherization radius,
see~\citet{SS73} Section IV).
Radiation pressure becomes most important in the inner parts
of the disc where one should take into account general relativity (GR) effects
to estimate rigorously the real limit when
standard disc accretion approximation is violated. 
In this paper, we attempt to estimate accurately where does a radiatively
efficient relativistic disc become super-critical and find that the Eddington
limit is scaled up by a factor of several due to geometrical and GR effects.  

The article is organized as follows. 
In Section~\ref{sec:thin}, we consider the
spherization radius and estimate the critical 
mass accretion rate in the thin-disc
approximation taking into account relativistic corrections.
In Section~\ref{sec:thick}, we consider the different effects connected to disc
thickness that appear to make the effective Eddington limit even higher and
besides change some of the properties of near-standard discs below the
Eddington limit. In Section~\ref{sec:disc}, we discuss the limitations of our
approach and make conclusions. 

\section{Thin disc approximation}\label{sec:thin}

Thin accretion disc as introduced by \citet{SS73} was historically the
first accretion disc model successfully applied to explain the observational
properties of the discs in X-ray binaries and active galactic nuclei (AGN). 
Fully relativistic thin accretion disc model was introduced by
\citet{NT73}. Vertical radiation flux in co-moving frame:
\begin{equation}\label{E:fluxNT}
F = \dfrac{3}{8\pi} \dfrac{GM\dot{M}}{R^3} \dfrac{\cal Q}{{\cal B}
  \sqrt{\cal C}} = \dfrac{3}{2} \dfrac{c^5}{\varkappa GM}
\dfrac{\mdot}{r^3} \dfrac{\cal Q}{{\cal B} \sqrt{\cal C}},
\end{equation}
where $r=\dfrac{R c^2}{GM}$ is normalized radius, $\mdot =
\dfrac{\dot{M}c^2}{L_{\rm Edd}} = \dfrac{\dot{M} \varkappa c}{4\pi GM}$ is
normalized mass accretion rate, and calligraphic letters correspond to the
coefficients introduced and calculated by \citet{NT73} and \citet{PT74}:
\begin{equation}
{\cal B}  = 1+\dfrac{a}{r^{3/2}},
\end{equation}
\begin{equation}
{\cal C} = 1-\dfrac{3}{r} + \dfrac{2a}{r^{3/2}},
\end{equation}
\begin{equation}
{\cal Q} = \dfrac{\cal B}{\sqrt{r {\cal C}}}
\left(\sqrt{r}-\sqrt{r_{in}}- \dfrac{3}{4}a\ln\dfrac{r}{r_{in}} - A_1-A_2-A_3 \right),
\end{equation}
where:
\begin{equation}
A_1 = \dfrac{3(x_1-a)^2}{x_1(x_1-x_2)(x_1-x_3)}\ln
\left(\dfrac{\sqrt{r}-x_1}{\sqrt{r_{\rm in}}-x_1}\right),
\end{equation}
$x_{1,2,3}$ are solutions of the equation $x^3-3x+2a=0$, and $A_2$ and $A_3$
are obtained from $A_1$ using cyclic permutation $1\to 2\to 3$.

\begin{figure}
 \centering
\includegraphics[width=\columnwidth]{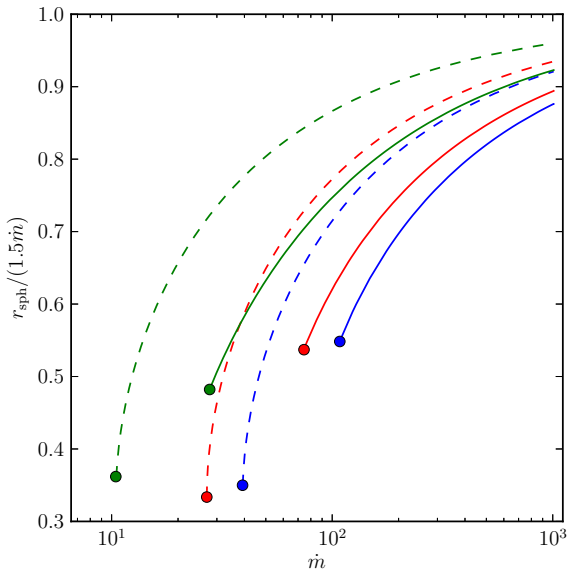}
\caption{Spherization radii calculated according to relativistic (solid lines)
  and non-relativistic (dotted) formulae. Thick dots mark the critical
  accretion rate for the relevant Kerr parameter value. The curves and dots
  were calculated for three different Kerr parameter values: $a=0$ (red,
  middle curves), $a=0.9$ (assuming co-rotation; blue, rightmost curves) and
  $a=-0.9$ (counter-rotation; green, leftmost curves). 
}
\label{fig:rsph}
\end{figure}

Radiation pressure is opposed by the vertical gravity that was first calculated
by \citet{RH95} in the $z/R \ll 1$ limit. Co-moving vertical gravity:
\begin{equation}\label{E:gz}
g_{\rm z} = \dfrac{GM z}{R^3} \dfrac{C_{\rm r}}{{\cal C}},
\end{equation}
where:
\begin{equation}
C_{\rm r} = 1-\dfrac{4a}{r^{3/2}} + \dfrac{3a^2}{r^2} .
\end{equation}
For a thin disc, vertical gravity is always proportional to the vertical
coordinate $z$. At some height radiation pressure becomes balanced by the
vertical gravity:
\begin{equation}
z= \dfrac{3}{2}\mdot  \dfrac{GM}{c^2} \dfrac{\cal Q}{{\cal B}  }
 \dfrac{\sqrt{\cal C}}{C_{\rm r}} .
\end{equation}
This quantity has the physical meaning of the height of a radiation pressure
supported accretion disc. It is also useful as an estimate for the so-called
spherization radius at which the thin disc approximation breaks down. Since
the vertical gravity approaches maximum at $z\sim R$, this puts also the local
Eddington limit for the mass accretion rate. A fully relativistic expression
for the spherization radius may be given in implicit form:
\begin{equation}\label{E:rsph}
r_{\rm sph} = \dfrac{3}{2} \mdot \cdot \dfrac{{\cal Q}(r_{\rm sph},a)}{{\cal B}(r_{\rm sph},a)}
\dfrac{\sqrt{\cal C}(r_{\rm sph},a)}{C_r(r_{\rm sph},a)} .
\end{equation}
This expression may be solved numerically for $r_{\rm sph}$.
The load of coefficients on the right-hand side approaches unity at large radii,
but the approach is slow enough to produce strong deviations from the
classical value of $r_{\rm sph}=\dfrac{3}{2}\mdot$ given by \citet{SS73}. These
coefficients incorporate both relativistic effects and the inner boundary
condition at $r=r_{\rm in}$ (at the last stable orbit radius). It is instructive to compare
expression~(\ref{E:rsph}) with its non-relativistic analogue taking into
account only the inner boundary condition \citep{superlens}:
\begin{equation}\label{E:rsphNR}
r_{\rm sph, NR} = \dfrac{3}{2} \mdot
\dfrac{4}{3}\cos^2\left(\dfrac{1}{3}\acos\left(-\dfrac{3\sqrt{r_{\rm in}}}{\sqrt{2\mdot}}\right)
\right) .
\end{equation}
In this approximation, spherization radius exists only if $\mdot>\mdot_{cr}=4.5r_{\rm in}$
In Fig.~\ref{fig:rsph}, we compare these two estimates. For accretion rates
$\mdot \lesssim 1000$, relativistic corrections alter the spherization radius
by about 20\%. In Fig.~\ref{fig:rsph}, solid curves correspond to the GR
case, and dashed to non-relativistic approximation. In both cases, middle
curves correspond to the Schwarzschild case ($a=0$) and the right and the left
curves -- to a moderate Kerr case with $a=0.9$ assuming co-and
counter-rotation, correspondingly. 

\begin{figure}
 \centering
\includegraphics[width=\columnwidth]{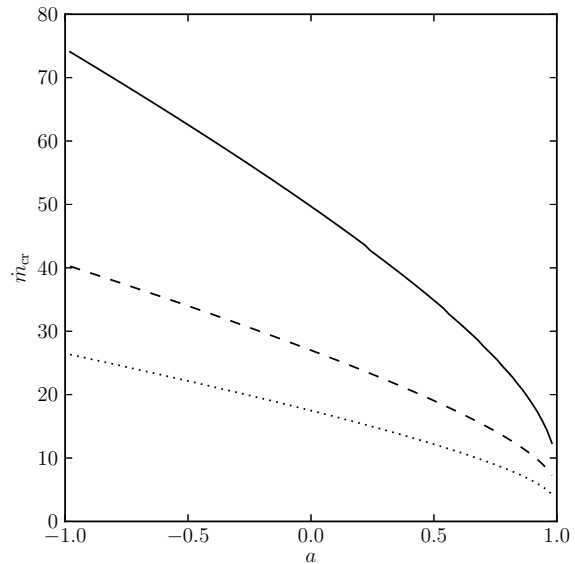}
\caption{Critical mass accretion rate as a function of Kerr parameter $a$
  (negative $a$ corresponds to counter-rotation). Dotted line is the critical
  accretion rate for spherically symmetric accretion ($\eta^{-1}(r_{\rm in},a)$),
  dashed line is the maximal mass accretion rate in Newtonian approximation, 
  and the solid line corresponds to the limit in full GR.
}
\label{fig:mcr}
\end{figure}

Equation~(\ref{E:rsph}) has no solutions if mass accretion rate is small
enough. If mass accretion rate is large enough, there are two solutions, the
larger of which is interpreted as the spherization radius. The minimal
\mdot\ value where the equation has a solution has the physical meaning of
critical mass accretion rate. While the spherization radius changes slightly due
to relativistic effects, the critical mass accretion rate defined this way
increases by about a factor of 3. In Fig.~\ref{fig:mcr}, we compare the
critical mass accretion rates defined straightforward as $L_{\rm Edd} /
\eta(a)c^2$, and non-relativistic and relativistic critical mass accretion
rates defined through local radiation pressure balance condition. Here, the
dotted line shows the limiting mass accretion rate in the spherically
symmetric case, equal to $\eta^{-1}(r_{\rm in},a)$. The dashed curve corresponds
to a thin Newtonian disc while the solid curve takes into account GR effects. 

\section{Corrections for disc thickness}\label{sec:thick}

\subsection{The issue of disc thickness}\label{sec:thick:issue}

As real accretion discs become thick near the Eddington limit, it is
instructive to check whether neglecting 
accretion disc thickness introduces any important corrections into
the vertical balance. Disc ``thickness'' is in fact a complex of several
physical effects connected to violation of the basic assumptions of the
thin-disc approximation, including:
\begin{enumerate}
\item geometrical thickness and, as a consequence, non-linear dependence of
  the vertical gravity on height,

\item non-Keplerian rotation and differential rotation with vertical
  coordinate that may lead to additional angular momentum transfer in vertical
  direction; this also affects dissipation and energy release inside the disc,

\item violation of local energy release approximation due to both advection
  and radial flux diffusion,

\item existence of radial and meridional velocity fields.

\end{enumerate}
Here, we will try to approximately account for the first, the third and the
partly the second correction.
We will assume
constant rotation frequency on cylindrical surfaces $\varpi = r\sin\theta =
const$, where $\theta$ is polar angle. This is close to the rotation law expected in barotropic case
where rotation frequency should be constant on the so-called von Zeipel
cylinders \citep{relvonzeipel}. 
In Appendix~\ref{sec:app:KvZ}, we show that
the shapes of von Zeipel cylinders are indeed very close to cylinders in
Boyer-Lindquist coordinates practically everywhere except inside the ergosphere.
Accretion disc enters the ergosphere only for very large rotation parameters
$a\gtrsim 0.94$. 
Outside the ergospheric region, for a fixed cylindrical coordinate $\varpi =
r\sin \theta$, rotation frequency always grows with height that implies faster
rotation (and, consequently, stronger effective gravity) at a surface for
constant spherical coordinate $r$. Hence using coordinate cylinders
$\varpi=const$ we underestimate the growth of the gravity with height. 

\subsection{Numerical simulations}\label{sec:thick:harm}

\begin{figure*}
 \centering
\includegraphics[width=1.1\textwidth]{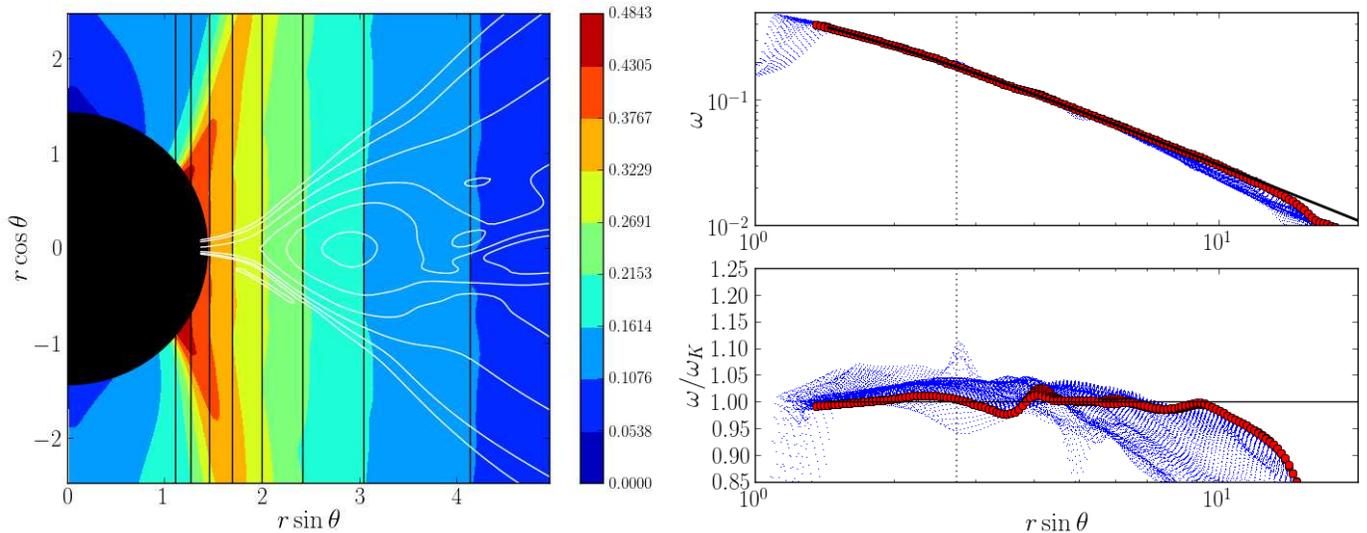}
\caption{ {\sc HARM2D} simulation snapshot ($t=5600\,GM/c^3$, B1h simulation
  from~\citet{azpaper}). Left panel: colour-coded is angular
  velocity $\Omega = u^\varphi / u^t$ in $c^3/GM$ units, 
black lines are vertical ($\varpi = const$) lines corresponding to the same
values of angular velocity as the boundaries between different colours/shades. 
Right panels: angular frequency slice through the
  equatorial plane (thick red points) and for all the points with polar angle
  cosines $|\mu| < 1/\sqrt{2}$; solid black line is Keplerian law $\omega =
  \left((r\sin \theta)^{3/2}+a\right)^{-1}$. Lower right panel shows relative
  deviations from the cylindrical Keplerian law. All the distances are in
  $GM/c^2$ units. Vertical dotted lines mark the radius of the last stable orbit.
}
\label{fig:harms}
\end{figure*}

Numerical simulations of thick MHD discs support the assumption
of Keplerian rotation in the inner parts of thick accretion discs and even
inside the last stable orbit \citep{penna10}. We have considered the
results of a thick disc MHD simulation from \citet{azpaper}, namely the run {\tt
  B1h} presented in this work. 
Simulation was performed using the numerical code {\sc HARM2D} that solves
ideal MHD equations in Kerr metric \citep{harm1, harm2}. The main goal of the
simulation was to form a moderately thin disc ($H/R \sim 0.1$) around a Kerr
($a=0.9$) black hole and consider the motion of the matter inside the inner
edge of the disc. {As the initial conditions, an exact
relativistic torus solution by~\citet{FM76}, section~IIIb, was used with small magnetic fields introduced that are}
subsequently amplified by MHD instabilities in the disc. Amplified magnetic
fields lead to outward angular momentum transfer and consequently to 
accretion from the torus through a moderately thick disc formed
during the first two thousand dynamical time-scales ($GM/c^3$) of the
simulation. {We also use the simulation run A50s presented in
  \citet{azpaper} that was calculated for $a=0$ and multi-loop initial 
magnetic field and
  uses smaller resolution. Main details of the simulations used in the present
  paper are given in Table~\ref{tab:harm}.

\begin{table}\centering
\caption{{\it HARM} model parameters, including the parameters of the initial
  Fishbone-Moncrief torus, $r_{max}$ (radius of maximal specific enthalpy) and $r_{in}$
  (inner radius) are also given. The time ranges (in $GM/c^3$) used for
  averaging are also given. }
\label{tab:harm}
\bigskip
\begin{tabular}{l|ccccp{1.0cm}p{1.5cm}}
\hline\\

Model & resolution & $a$  & $r_{\rm max}$ & $r_{\rm in}$ &  number of loops & time
range \\

\hline\\

B1h & 192$\times$320 & 0.9 & 8.4   & 4.8  &  1 & 2000-5700\\ 
A50s & 128$\times$192 &  0      &  19.7   & 12.8  &  50 & 10000-15000 \\
D1  & 128$\times$192 & 0.99   & 10  & 4 & 1 &  2000-5700 \\
\hline
\end{tabular}
\end{table}

A thicker torus in the model of \citet{FM76} is also more extended in radial
direction. As the result, it may become impossible to maintain the boundary
condition at the outer boundary, and the simulation becomes unstable. A way
around is to consider a black hole with a higher Kerr parameter. More rapid
rotation
allows to produce a thicker disc without making the initial torus model
extended beyond the simulation domain.
Apart from the models described in \citet{azpaper}, we have calculated a
similar model with a higher Kerr parameter, $a=0.99$.  Absence of
heating and cooling sources
in {\sc HARM2D} makes it difficult to simulate very thick as well as very thin
discs. While the initial torus has $H/R \sim 0.5$, the accretion disc forms from its
inner cusp that has lower geometrical thickness. 
For the B1h model, the relative thickness is $H/R \simeq 0.1$ as
measured by Gaussian fitting, while for the model with $a=0.99$ that we will
hereafter dub as D1, the thickness was around $0.2$. In fact, all the
equilibrium torus models, however thick, have smaller thicknesses near the
inner rim (see, for example,~\citet{penna13}, Fig. 3). 
 
Surprisingly, in all our models,} not only the lines of constant
angular velocity are close to cylinders $\varpi=r\sin\theta=const$ but the rotation law
between $r\simeq 2$ and $r\simeq 10$ conforms with Keplerian rotation with
about 3$\div$4 per cent accuracy (relative standard deviation from the
Keplerian law) for $\theta \gtrsim 45^\circ$.  For larger
distances from the equatorial plane, rotation becomes strongly sub-Keplerian. 
On the other hand, in the equatorial plane, Keplerian rotation holds to about 1
per cent. 
In Fig.~\ref{fig:harms} (left panel), 
lines of constant angular frequency ($\Omega = u^\varphi / u^t$) are shown and
compared to the Keplerian rotation law ($\Omega_{\rm K} =
\left(\varpi^{3/2}+a\right)$), and the lines of constant angular
frequency are shown to conform well with the cylinders $\varpi =
const$. Keplerian rotation holds even inside the last stable
orbit. 

{

\begin{figure*}
 \centering
\includegraphics[width=1.1\textwidth]{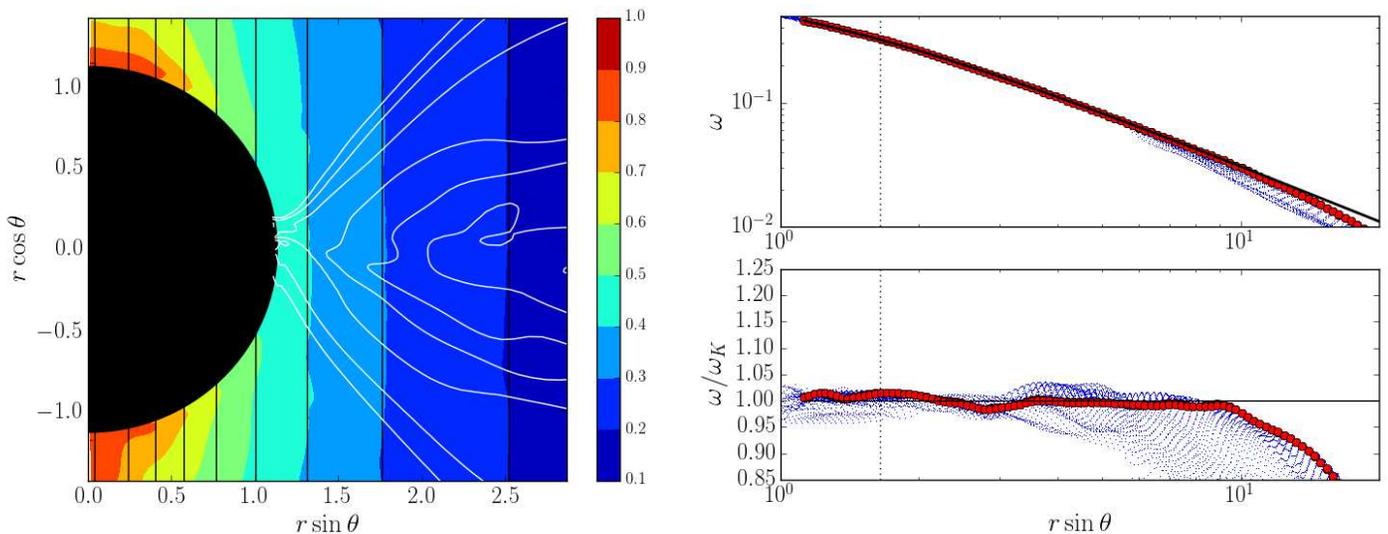}
\caption{ {\sc HARM2D} simulation snapshot ($t=5400\,GM/c^3$, D1 simulation
  with $a=0.99$). Notation is the same as in the previous figure.
}
\label{fig:harmsD1}
\end{figure*}

Our results are at odds with some of the comprehensive thick-disc simulations
such as by \citet{supersadowski} that, for a similar set-up, find rotation
sub-Keplerian by not less than 15 per cent, both outside and inside the last
stable orbit (see Fig.~11 in this work). On the other hand, \citet{shafee},
for instance, find the angular momentum close to Keplerian just outside the
last stable orbit. The overall behaviour of the net angular momentum in the
simulations by \citet{shafee} conforms
well with the conventional picture of hydrodynamic transonic accretion: 
nearly Keplerian rotation outside the last stable orbit, Keplerian rotation
at the sonic surface and roughly constant angular momentum ($u_\varphi$) in
the supersonic region. On the other hand, \citet{supersadowski} find the net
angular momentum rapidly decreasing inside the last stable orbit. 

The reason for this discrepancy lies probably in the diverse structure of
magnetic field formed during disc accretion with frozen-in magnetic fields
\citep{penna10}. The
initial conditions include small seed magnetic field subsequently
amplified by MHD instabilities. Amplified poloidal magnetic fields are then
advected towards the black hole and contribute to the accumulated magnetic
flux through the black hole horizon. 
This magnetic flux grows gradually with time that leads to secular evolution
of the deviations from Keplerian rotation. In simulations B1h and D1, we
assumed the initial
magnetic field potential scaling with the rest-mass density as $A_\varphi \propto
\rho$. However, in \citet{supersadowski}, the magnetic field obeys a different,
more complicated law making it stronger further from the black hole. This
leads to a gradual increase in magnetization of the accreting matter even during
quasi-stationary accretion \citep[see also Fig.~6 of][]{supersadowski}.

We conclude that the rotational structure of the inner disc and especially the
flow inside the last stable orbit is strongly affected by the magnetic fields
and especially by the magnetic field flux accumulated inside the disc. Our
results should work as long as the large-scale magnetic field is dynamically
unimportant. For the time span of the B1h simulation, rotation
law changes very weakly after the establishment of quasi-stationary
accretion at $t\sim 2000\,GM/c^3$, but simulations containing more complicated
magnetic fields like A50s show irregular variations due to changes in magnetic
field strength and geometry (see Fig.~\ref{fig:levol}). Angular momenta given
in this figure are density-weighed vertically-averaged values (as
in~\citet{supersadowski}). 

\begin{figure*}
\includegraphics[width=1.1\textwidth]{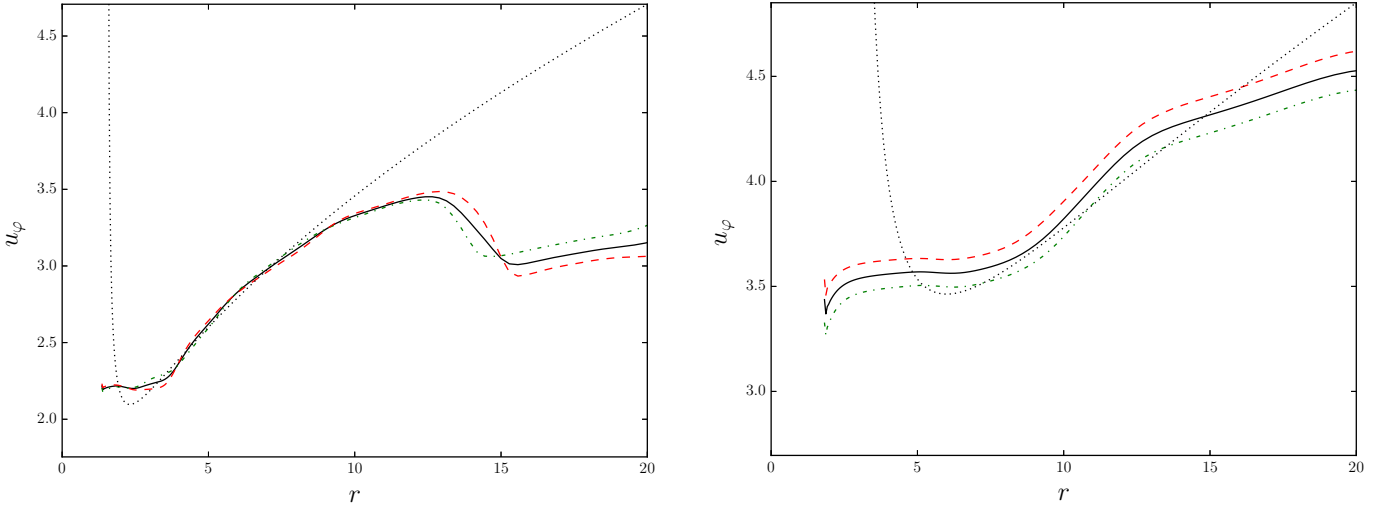}
\caption{ Radial profiles of density-weighed specific angular momenta for two simulations
  B1h (left) and A50s (right). Black solid curves are averages for all the frames of
  the quasi-stationary phase, green dot-dashed and red dashed curves show the averages
  calculated for the first and second halves of the period of
  time that we
  interpret as quasi-stationary accretion ($2000\div 5700\,GM/c^3$ for B1h,
  $10\,000\div 15\,000\,GM/c^3$ for A50s).
}
\label{fig:levol}
\end{figure*}

In our simulations,} rotation frequency is everywhere in the thick disc close
to the following ``cylindrical Keplerian'' law:
\begin{equation}
\Omega(r, \theta) = \dfrac{1}{\varpi^{3/2}+a} .
\end{equation}
Estimating the two-dimensional rotational velocity field is crucial for the
vertical structure of the disc since the vertical structure of the disc is
determined by the deviations from the balance between inertial and
gravitational forces, the former dependent upon the rotation law. There is no clear
distinction between gravity and inertia in general relativity, and the
vertical hydrostatic balance in accretion disc may be alternatively viewed as
a balance between pressure gradients in the disc and the centrifugal force
\citep{ALP97}.

\subsection{Corrections for vertical gravity}\label{sec:thick:vert}

Let us use spherical coordinates and
consider the balance of radiation pressure and inertial forces (including
gravity) along the polar angle $\theta$. It is more convenient to use $\mu = \cos
\theta$ instead of the angle itself or the vertical distance $z$. Relevant
metric coefficient is 
\begin{equation}
g_{\mu\mu} = g_{\theta\theta}
\left(\dfrac{d\theta}{d\mu}\right)^2 = \rho^2/(1-\mu^2).
\end{equation}

Then, considering a sphere of
$r=const$, one can express the ``vertical'' free-fall acceleration as:
\begin{equation}
g_\mu = -\Omega_\mu^2\mu = \Gamma_{i \mu k} u^i u^k,
\end{equation}
where $i$ and $k$ may be equal $t$ or $\varphi$, $u^\varphi = \Omega u^t$, and
$u^t$ is recovered from normalization condition $u_i u^i =-1$ (we neglect the
radial and meridional velocities here). Vertical gravity should
be compared to the coordinate flux component $F_\mu = F \sqrt{g_{\mu \mu}}$
multiplied by $\varkappa /c$ (see below Section~\ref{sec:thick:adv}). All the
relevant non-zero Christoffel symbols are $\Gamma_{i \mu k} = \dfrac{1}{2}
g_{ik, \mu}$:
\begin{equation}
\Gamma_{t\mu t} = -\dfrac{2a^2r}{\rho^4}\mu ,
\end{equation}
\begin{equation}
\Gamma_{t\mu \varphi} = \Gamma_{\varphi \mu t} =
\dfrac{2ar\left(r^2+a^2\right)}{\rho^4}\mu ,
\end{equation}
\begin{equation}
\begin{array}{l}
\Gamma_{\varphi \mu \varphi} = -\dfrac{1}{\rho^4} \left( a^2(1-\mu^2)
\Sigma^2+\rho^2\Sigma^2 - a^2\Delta (1-\mu^2)\right)\mu=\\
=-\dfrac{r^6+a^2r^4+4a^2r^3+2a^4r+a^2\Delta\rho^2\mu^2}{\rho^4} \mu .\\
\end{array}
\end{equation}
Substituting these expressions in the expression for $g_\mu$, up to second
order in $\mu$:
\begin{equation}\label{E:mu:gz}
\Omega_\mu^2 = -\dfrac{1}{r} \left( \dfrac{C_r}{\cal C} + \dfrac{3}{2}
\dfrac{\zeta(r,a)}{{\cal C}^2 {\cal B}} \mu^2\right),
\end{equation}
where:
\begin{equation}
\begin{array}{l}
\zeta(r,a) = 1-\dfrac{8}{3} \dfrac{1}{r} + \dfrac{a^2}{r^2} + \dfrac{14}{3}
\dfrac{a^2}{r^{5/2}} - \dfrac{23}{3} \dfrac{a^2}{r^3}
+\dfrac{16}{3}\dfrac{a^3}{r^{7/2}}  + \\
\qquad{} + \dfrac{2}{3}
\dfrac{a^2(7-4a^2)}{r^4} -
 12 \dfrac{a^3}{r^{9/2}} + 17 \dfrac{a^4}{r^5} - \\
\qquad{} -\dfrac{8}{3} \dfrac{a^3 (1+3a^2)}{r^{11/2}} - \dfrac{16}{3} \dfrac{a^4}{r^6} +
\dfrac{38}{3} \dfrac{a^5}{r^{13/2}}-\dfrac{16}{3} \dfrac{a^6}{r^7} .\\
\end{array}
\end{equation}

In such an approach, for most radii and Kerr parameter values, 
the effective vertical frequency
becomes higher away from the equatorial plane if the spherical radial
coordinate is held fixed. This effect arises from the
assumption $\Omega=\Omega(r\sin\theta)$ that leads to effectively
super-Keplerian rotation near the disc surface. If one considers a more
realistic rotation law with $\Omega$ constant on von Zeipel cylinders, the
effect becomes even stronger, except the most rapidly rotating black holes
($a\gtrsim 0.95$)
where the inner parts of the disc exist inside the ergosphere, where the
shapes of von Zeipel cylinders change considerably (see
Appendix~\ref{sec:app:KvZ}). In our assumptions, $\zeta$ and thus the whole
next-order correction is always positive. We also assume that the maximal
possible thickness of the disc is $\mu_{\rm max} = 1/\sqrt{2}$. 

%

\begin{figure*}
 \centering
\includegraphics[width=\textwidth]{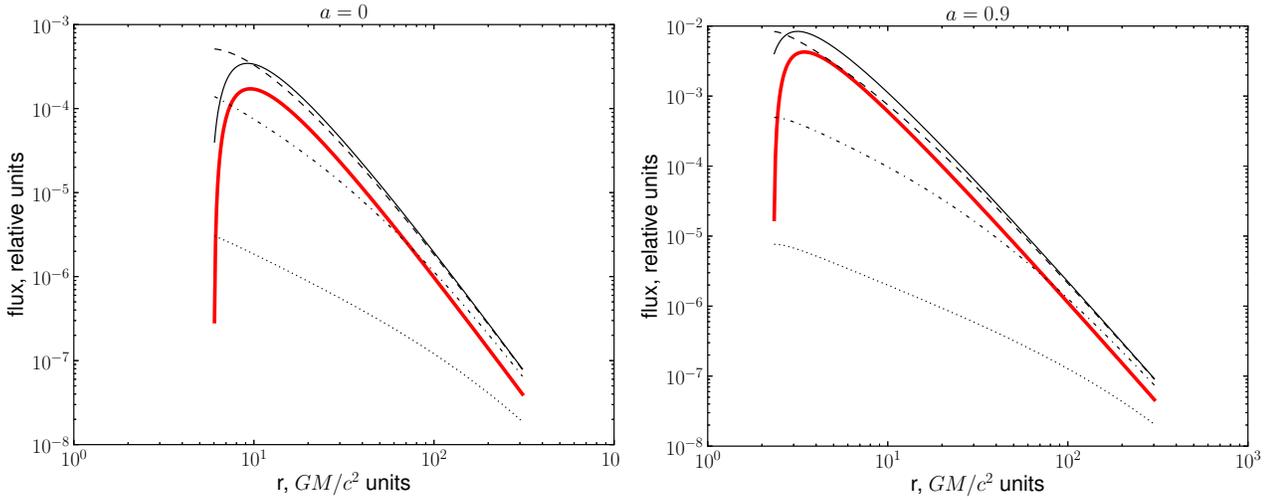}
\caption{Flux dependence on radius for different mass accretion rates. Fluxes
  are divided by the mass accretion rate hence the highest fluxes correspond
  to the smallest \mdot\ where advection is smaller. The thick red curve shows the
  standard accretion disc model, thin black curves correspond to $\mdot = 1$,
  $10$, $100$ and $1000$ (solid, dashed, dash-dotted and dotted lines,
  respectively). Left and right panels show the cases of $a=0$ and $a=0.9$. 
}
\label{fig:fluns}
\end{figure*}

\subsection{Advection}\label{sec:thick:adv}

Local energy flux in an axisymmetric disc consists of diffusion and advection
terms, the latter directed along the local poloidal velocity. If the advection
term is small compared to vertical diffusion, its overall effect is to shift
downstream the thermal energy generated in the disc.
Weak advection in a moderately thick disc 
may be approximately taken into account by introducing a ``delayed
radius'' assumption that the flux generated at radial distance $r_0$ will be
then emitted at a smaller radius $r=r_0-\delta r(r_0)$. For our purposes, this
approach is better than assuming a fixed fraction of liberated energy stored
in the thermal energy of the disc as is it often done when considering discs
with advection \citep{NY95, slim}, since it does
not include unknown free parameters and can account for advection dependence
on radius. 
Apart from the pure flux advection, one should take into account that the
frame where the flux is compared to the vertical gravity is different from the
frame (or frames) where the energy is generated that we would further identify
with the co-moving equatorial frame in the disc. 
Radial shift $\delta r$ is a result of interplay between vertical energy
diffusion and radial drift. 
\begin{equation}
\delta r \simeq v_{\rm r}  t_{\rm diff},
\end{equation}
where $v_r$ is the radial velocity measured in the reference frame co-moving
with the disc, $t_{\rm diff} = H \tau_{\rm v}$, where $H=r\mu_{\rm max}$, and
$\tau_{\rm v} = \dfrac{1}{2}\varkappa \Sigma$ is the vertical optical depth between the
equatorial plane and the disc surface. 
\begin{equation}\label{E:dr}
\delta r \simeq \dfrac{\mu_{\rm max}\mdot}{2}
\end{equation}
The main simplification is that the flux generated at the radial coordinate
$r_0$ is emitted in a narrow radius interval near $r$. A reasonable assumption
is also to assume all the flux generated near the equatorial plane and then
conserved. More precisely, let us consider the energy flux due to radiation
$T^k_{\rm t} = u_{\rm t} F^k + u^k F_{\rm t}$ and 
integrate it upon a $\mu=\mu_{\rm max}$ surface just above the surface
of the disc. Conservation of the energy flux implies:
\begin{equation}
T^k_t \sqrt{-g}dS_k = const,
\end{equation}
for a given flux tube with cross-section given by the four-vector
$dS_k$. Setting the cross-section tangent to the $\theta=const$ surface implies
\begin{equation}
u_t F^\mu \sqrt{-g} dS_\mu = const,
\end{equation}
or
\begin{equation}
u_t^{\rm (s)} \sqrt{-g_{\rm (s)}} \dfrac{F_{\rm (s)}}{\sqrt{g_{\mu\mu}^{\rm
      (s)}}} dr_{\rm (s)} d\varphi = u_t^{\rm (d)} \sqrt{-g_{\rm (d)}}
\dfrac{F_{\rm (d)}}{\sqrt{g_{\mu\mu}^{\rm (d)}}} dr_{\rm (d)} d\varphi,
\end{equation}
where quantities with `(d)' index refer to the equatorial plane, and with
`(s)' index -- to the disc surface, $g = -\rho^4$ is metric
determinant (hence $g_{\rm (d)} = -r_0^4$). Since $dr_{\rm (d)} = dr_{\rm
  (s)}$ (see above expression (\ref{E:dr}) for $\delta r$), area
distortion is connected only to the
difference in $\sqrt{-g}$ factors, incorporating relativistic effects and
spherical geometry. 
Co-moving flux at the surface of the thick disc:
\begin{equation}\label{E:fs}
F_{\rm (s)} = \dfrac{r_{(d)}}{\rho_{(s)}\sin \theta} \dfrac{u_t^{\rm
    (d)}}{u_t^{\rm (s)}}
F_{\rm (d)}. 
\end{equation}
All the quantities with a `(d)' index on the right-hand side should be
evaluated at $r_0$, and $\theta$ is the half-opening angle of the
disc. Eddington limit in polar angle direction is written as:
\begin{equation}\label{E:eddthick}
\dfrac{\varkappa}{c} F_{\rm (s)} \sqrt{g_{\mu\mu}} = \Omega^2_\mu \mu,
\end{equation}
where $\Omega_\mu$ is given by expression~(\ref{E:mu:gz}), and flux $F_{\rm (s)}$ by expression~(\ref{E:fs}). 

\subsection{Effects of disc thickness upon the local Eddington limit}\label{sec:thick:edd}

The effects of non-locality (including advection) are ambiguous: flux increases
due to geometrical effects, and decreases due to decreasing dependence $F(r)$ 
everywhere except the innermost parts where the disc releases more energy than
a standard disc.
In Fig.~\ref{fig:fluns}, flux dependence on radius is shown taking into account
advection and the geometrical effects described in
Section~\ref{sec:thick:adv}. Note that
for very high mass accretion rates ($\mdot \gtrsim 10^3$) advection is able to
prevent the disc from surpassing the local Eddington limit. 
Our approximation does not reproduce this regime but it can definitely appear when advection
becomes strong and essentially non-linear. This advection-dominated regime may
be loosely identified with the slim-disc solution \citep{abram88,slim}. 

\begin{figure*}
 \centering
\includegraphics[width=\textwidth]{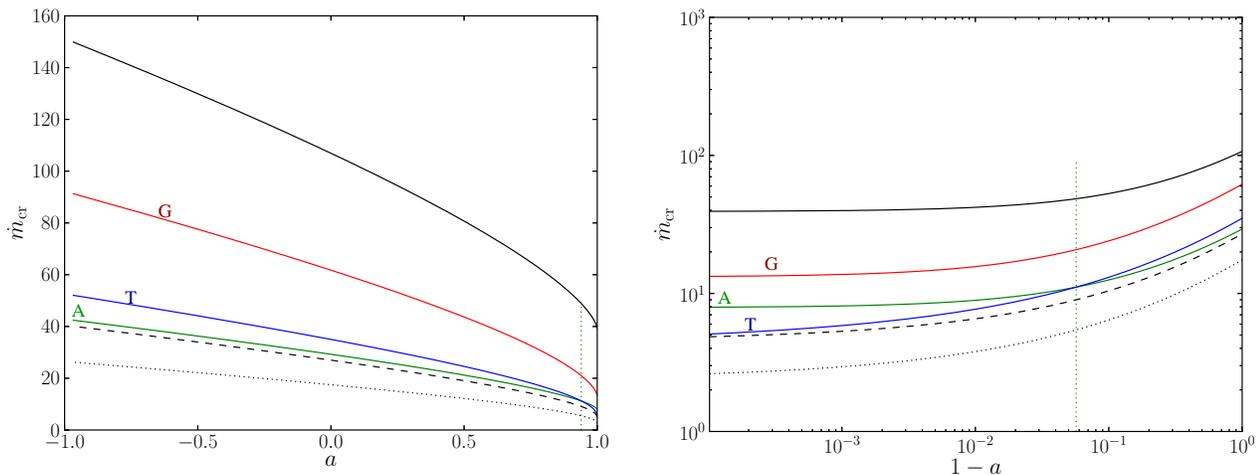}
\caption{Critical mass accretion rate as a function of Kerr parameter. Left
  and right panels differ only in abscissa: in the left, it is linear in $a$,
  in the right, it is made logarithmic in $1-a$ to resolve the behaviour of
  limiting mass accretion rates at high mass accretion rates. 
  Dotted and dashed black curves are identical to those in
  Fig.~\ref{fig:mcr}, corresponding to spherical and disc non-relativistic
  Eddington limit, correspondingly. The solid blue line (also marked as ``T'') corresponds to the
  local thin relativistic disc Eddington limit. It goes somewhat lower than
  the solid line in Fig.~\ref{fig:mcr} because the accretion disc thickness
  is limited by $\mu=1/\sqrt{2}$ rather than $\mu=1$. Green curve (marked by ``A'') takes into
  account advection only, and red curve (marked by ``G'') only second-order corrections to
  vertical gravity. Solid black curve accounts for both effects. Vertical
  dotted line marks the change of sign in $\zeta$ correction at the inner rim
  of the disc. 
}
\label{fig:mlimses}
\end{figure*}

In Fig.~\ref{fig:mlimses}, critical mass accretion rates calculated with
and without corrections for disc thickness (including advection) are shown. As in
Fig.~\ref{fig:mcr}, dotted line corresponds to the spherically symmetric
case, and dashed line to the thin non-relativistic disc. Blue line (marked
with a ``T'' in the figure) shows the
thin relativistic disc case considered in Section~\ref{sec:thin}. However, the
limiting disc thickness is taken to be $(H/R)_{\rm max} =1/\sqrt{2}$ (since
$R$ is interpreted as spherical rather than cylindrical coordinate; this is
also close to the maximal polar angle
cosine where cylindrical rotation holds in the {\sc HARM2D} simulation we
mention above) that decreases the critical rate. Red (subscribed with a ``G''
in the figure) and
green (marked ``A'') curves show the 
effects of gravity and advection taken separately, and
the solid black curve is calculated taking into account both effects. The
critical mass accretion rates are shown as functions of Kerr parameter
$a$. Same critical mass accretion rates are also shown as functions of $1-a$
in logarithmical scale that allows to show in detail the limit of rapid black
hole rotation.

Evidently, corrections
to vertical gravity increase the value of the Eddington limit, at least for
$a\lesssim 0.96$. However, the resulting values of the critical mass accretion
rate are not much larger than the thin-disc estimate. 
While an account of advection and pure spherical geometry effects described in
Section~\ref{sec:thick:adv} in fact {\it decreases} the local Eddington limit
because the local flux in the inner parts of the disc increases, stronger
vertical gravity acts in the opposite direction. Including both effects simultaneously
increases the local flux limit even more than gravity alone because at larger
$\mdot$,
advection effects decrease the local flux. The minimal possible spherization
radius also becomes considerably larger because in
the inner parts of the accretion disc, advection becomes too strong. 
Summarizing, we conclude that different effects connected to disc thickness
produce corrections of different signs, but the general result holds and
becomes even stronger.

\subsection{Zone $a$ disc thickness}\label{sec:thick:athick}

\begin{figure*}
 \centering
\includegraphics[width=\textwidth]{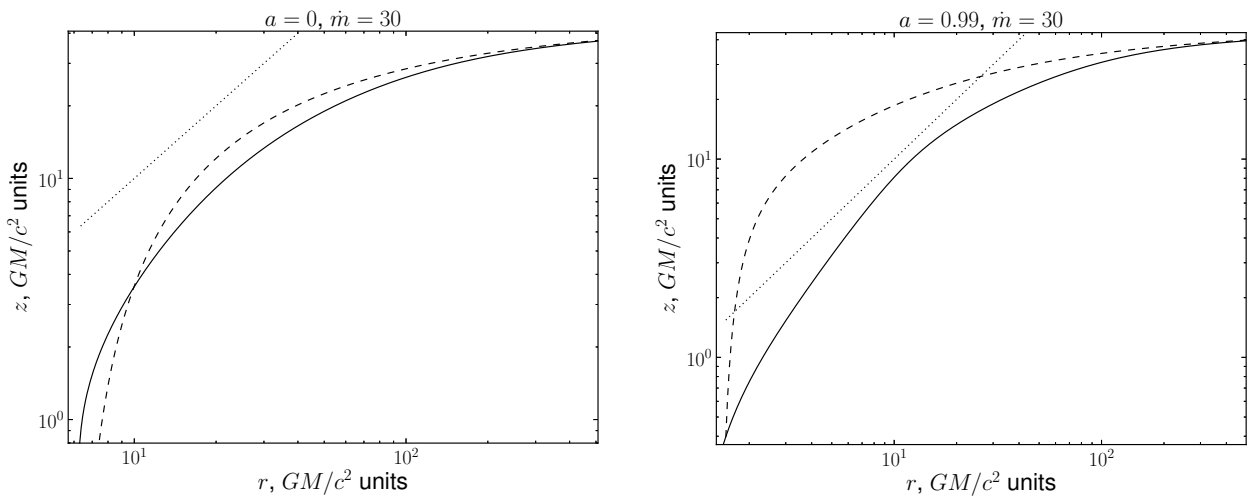}
\caption{ Disc vertical scales calculated taking into account corrections for
  vertical gravity and advection (solid lines). Dashed lines are standard disc
  zone $a$ thicknesses for identical parameters. Dotted straight lines
  correspond to $z=r$. 
}
\label{fig:hths}
\end{figure*}

Physical meaning of condition (\ref{E:eddthick}) is vertical balance between
radiation and gravity, it can also be solved for $\mu$ to estimate the
vertical scale of the disc. In Fig.~\ref{fig:hths}, we compare the 
vertical scales of the discs for $\mdot=30$ and different Kerr parameters. We
find that disc thickness corrections decrease the vertical scale in the
hottest and thickest parts of the disc while its inner rim attains finite
thickness. At high mass accretion rates, the thickest part of the disc shifts
to larger distances, and the surface inside the thickest part approaches a
roughly paraboloidal or conical shape.

For disc thicknesses, corrections for advection were calculated using
expression (\ref{E:dr}) for radial shift. 
In reality the radial advection correction to the pressure inside the disc
should be smaller because the thermal energy released inside the disc has an
immediate effect upon the local pressure but delayed effect upon the flux
coming out from the disc surface.
 Future calculations of vertical disc structure should take this into
account together with radial diffusion of radiation and other effects
connected to the two-dimensional nature of thick accretion discs. The general
effect upon the vertical scale is smearing the abrupt decrease in disc
thickness near the inner radius. The shape of the disc surface generally
becomes closer to $h/r = const$. 

\subsection{Efficiency}\label{sec:thick:eff}

Increasing mass accretion rate limit does not necessarily mean larger
luminosity because some part of the energy dissipated inside the disc is
advected inwards. Since $\mdot \gg 1$, the energy is trapped in the accreting
flow even inside the last stable orbit. Indeed, the vertical optical depth in
the nearly free-fall region inside the last stable orbit is
\begin{equation}
\tau_z = \varkappa \int_0^z (-u_i k^i) \rho d\lambda,
\end{equation}
where $\lambda$ is affine parameter increasing along the photon trajectory and
$k^i = dx^i / d\lambda$. Mass conservation implies $\Sigma = \dot{M} / 4\pi R
u^r$. For a vertical trajectory, $dr=0$ and $d\theta=0$, and hence, if $u^z=0$:
\begin{equation}
\tau_z = \varkappa \Sigma  E = \dfrac{E \mdot }{2r u^r},
\end{equation}
where $E=-u_t$ is energy-at-infinity for a unit-mass particle in the flow. 
The locally measured diffusion time is $t_{\rm diff} \sim H
\tau_z/c$. Radiation diffuses in vertical direction, but in radial direction
it is advected at the local dynamical time scale $t_{\rm dyn} \simeq c(r_{\rm
  ISCO} - r_{\rm hor})$, where $r_{\rm ISCO}$ and $r_{\rm hor}$ are the radii
of the last stable orbit and the event horizon of the black hole, and the
infall velocity was set to $c$. Relative importance of the two effects is
given by the time ratio
\begin{equation}
\dfrac{t_{\rm diff}}{t_{\rm dyn}} \simeq \dfrac{E \mdot}{2 (r_{\rm ISCO} -
  r_{\rm hor})} \dfrac{H}{R} \sim \mdot.
\end{equation}
This means that the falling flow inside the last stable orbit starting
from moderate accretion rates $\mdot \sim 1$ is indeed optically thick {to
  electron scattering},
especially for higher rotation parameters when the radial free-fall time is
small. {Optical depths to different absorption processes are likely
  smaller than unity but are sensitive to clumping (all the opacities that
  scale with density squared) or to magnetic field strength (synchrotron
  absorption) and hence are more difficult to constrain. Below in
  section~\ref{sec:thick:seds} we estimate the effective optical depth to
  free-free absorption.}

Since the flow is optically thick one can apply the formalism we used in
Section~\ref{sec:thick:adv}. The energy released in the disc will be shifted
downstream by:
\begin{equation}
\Delta r \simeq u^r t_{\rm diff} \simeq \dfrac{1}{2}\mu_0 \mdot E.
\end{equation}
This allows not only to estimate the overall efficiency but also the
corrections to the spectral energy distribution that we will present in
Section~\ref{sec:thick:seds}. 

\begin{figure}
 \centering
\includegraphics[width=1.1\columnwidth]{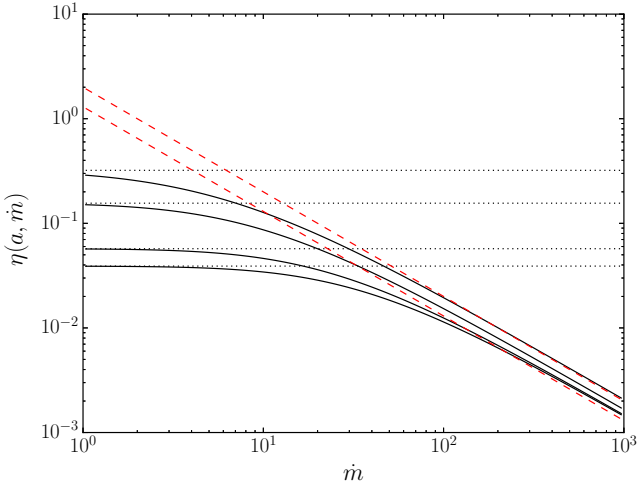}
\caption{ Efficiency dependence on mass accretion rate
  calculated in delayed radius approximation for different Kerr parameters
  ($a=-0.9$, 0, 0.9 and 0.998, shown by black solid lines, from bottom to
  top). Dotted lines are thin-disc
  efficiencies for corresponding Kerr parameters. Red dashed straight lines are
  $1.3/\mdot$ and $2/\mdot$. }
\label{fig:etadrop}
\end{figure}

\begin{figure}
 \centering
\includegraphics[width=1.1\columnwidth]{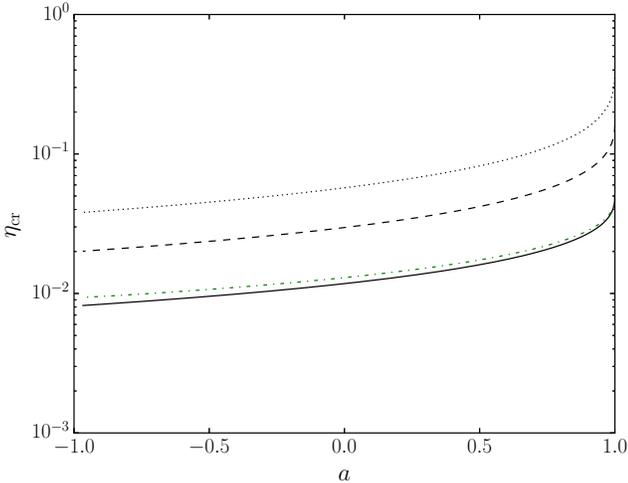}
\caption{ Efficiencies for a thin disc (dotted), thin disc with advection
  (dashed) and thick disc with advection (solid black and dot-dashed green lines). 
The green curve also takes into
  account the radiation escaping from inside the last stable orbit. 
  The three latter curves were calculated
  for corresponding critical mass accretion rates. 
}
\label{fig:etas}
\end{figure}

Accretion disc luminosity: 
\begin{equation}
\begin{array}{l}
L_{max} \simeq L_{\rm Edd}  \eta(a, \mdot_{\rm cr}) \mdot_{\rm cr}(a) \simeq \\
\qquad{} \simeq 1.4\times 10^{38}
\left(\dfrac{M}{\Msun}\right) \eta(a, \mdot_{\rm cr}) \mdot_{\rm cr}(a) \ergl.\\
\end{array}
\end{equation}
For a thin disc, (see, for instance, \citet{ST} Section 12.7):
\begin{equation}
\eta(a, \mdot=0) = 1-\sqrt{1-\dfrac{2}{3r_{\rm in}}}
\end{equation}
In the general case, $\eta = 1-E$ where energy $E$ is calculated at the inner
visible edge of the disc. Considered in the delayed radius formalism, $E$
should be taken at a radius larger than the last stable orbit radius. 
For our model thick disc with advection, the large optical depth of the
falling flow implies
\begin{equation}
\eta(a, \mdot) = 1-E(r_{\rm hor}+\delta r).
\end{equation}
Assuming Keplerian rotation allows to approximate the energy as
(cf.~\citet{PT74}; our estimate hear does not account for radial motions that
should be slow as long as Keplerian rotation holds):
\begin{equation}
E = \dfrac{{\cal G}}{\sqrt{\cal C}},
\end{equation}
where:
\begin{equation}
{\cal G} = 1-\dfrac{2}{r} + \dfrac{a}{r^{3/2}}.
\end{equation}
Efficiency starts decreasing significantly from $\mdot \sim 10$ and then
approaches a power-law asymptotic $C/\mdot$ with $C$ depending upon Kerr
parameter. For high mass accretion rates, this
implies a constant luminosity of $CL_{\rm Edd}$ with $C$ between approximately
1.3 and 2
(see Fig.~\ref{fig:etadrop}). 
When mass accretion rate reaches the critical value and the upper layers of
the disc become unbound, efficiency becomes smaller than for the standard thin
disc by a factor of about a factor of 5 (see Fig.~\ref{fig:etas}). 
Some, not very
large, part of this luminosity is emitted from the disc extention inside the
last stable orbit. As advection enters non-linear regime, and the disc starts
losing its matter, a logarithmical correction to this quantity should appear
that we will discuss in more detail in Section~\ref{sec:disc}. 

\subsection{Spectral energy distribution}\label{sec:thick:seds}

\begin{figure}
 \centering
\includegraphics[width=1.1\columnwidth]{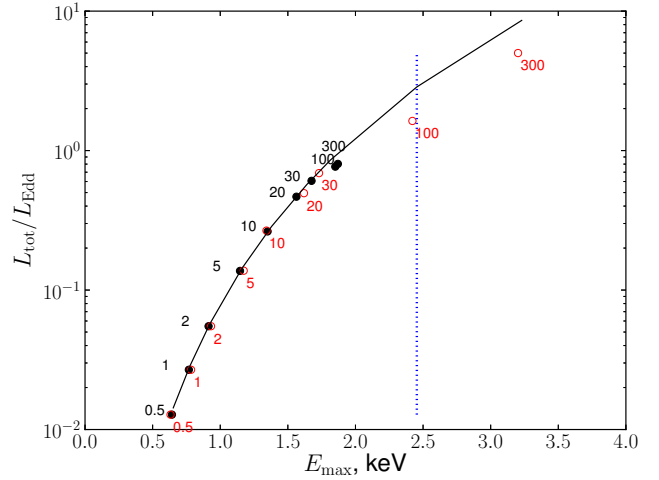}
\caption{ Temperature-luminosity plot (energy of maximal $\nu L_\nu$ in the
  spectrum was used as a measure of temperature) for black hole accretion discs with
  different mass accretion rates between 0.1 and 500 spaced evenly in
  logarithmic space. Dimensionless mass accretion rates are given as black and
  red numbers. Black circles are for
  a thick advection-modified disc, red open circles for a thin standard
  disc. Solid curve corresponds to the approximate relation for a standard
  accretion disc where $L\propto E_{\rm max}^4$. 
Dotted curve is an estimate for the maximal effective temperature
  expected in discs with outflows. 
}
\label{fig:scythe}
\end{figure}

As one can see from previous sections, including thickness effects makes
the flux generally higher for intermediate mass accretion rates and lower for
$\mdot\gtrsim 100$. Thus, the spectral energy distribution first becomes
hotter, then cooler with mass accretion rate.  
We have considered spectral energy distribution of a
multi-blackbody disc for a fixed configuration (Schwarzschild black hole
viewed at 30$^\circ$ inclination). Public code {\sc geokerr} (see~\citet{geokerr} for
description) was used to compute the
geodesics, and the observed intensity was re-calculated using the conservation
of energy and angular momentum along the trajectories. Details of intensity
transformation may be found e. g. in section~3.3 of \citet{paperI}. 
Black hole mass was set to $M=10\Msun$.  For
different mass accretion rates, spectral energy distributions were
calculated with and without considering thickness effects. For every spectrum,
we calculated integral luminosity and position of the maximum of $\nu F_\nu$
(flux per logarithmic frequency interval). The resulting
luminosity-temperature plot is given in Fig.~\ref{fig:scythe}. 

Because in our model the disc retains its multi-blackbody nature, it does not
shift significantly from the standard disc luminosity-temperature
relation. Most of the luminosity is emitted within the area comparable to the
area subtended by the last stable orbit hence the luminosity is always proportional
to the local flux in the inner parts of the disc and $L \propto T_{\rm max}^4
\sim E_{\rm max}^4$, where $E_{\rm max}$ is the energy where the spectral
energy distribution has the maximum. In Fig.~\ref{fig:scythe}, solid curve
is the approximate temperature in the disc without correction factors:
\begin{equation}
T_{\rm max} = 5.29 \left(\dfrac{\mdot}{M_1}\right)^{1/4} r_{\rm in}^{-3/4} {\rm
keV}
\end{equation}
For the set of parameters we use here, $T_{\rm max} \simeq 2 \mdot^{1/4}$keV. 

Strong differences (more than 10 per cent for luminosity, 5 per cent for the maximal
energy) in the spectral energy distributions appear above $\mdot\simeq 25$ and rapidly
increase with the mass accretion rate. This is considerably lower than the
$\mdot_{cr} \sim 100$ Eddington limit. 

{
Here we assume that all the energy is released in the disc and that all the
radiation is thermalized. First assumption is not strictly true because viscous stresses
should be present everywhere throughout the transonic flow (see simulations by
\citet{shafee,penna}). The energy released in the inner parts of the flow
contributes to the
temperature of the gas and ultimately to comptonization of the soft radiation
of the disc. We do not expect the optical depth of the flow to true
absorption to be significantly large. 
Following \citet{RL}, one can write down the free-free absorption
cross-section for fully ionized hydrogen plasma as
\begin{equation}
\sigma_{ff} \simeq \displaystyle \frac{4e^6}{3m_e c} \sqrt{\frac{2\pi}{3m kT}}
n_{\rm e} \frac{1-e^{-h\nu / kT}}{\nu^3},
\end{equation}
where $n_e$ is electron density that may be substituted as 
\begin{equation}
n_{\rm e} \simeq \dfrac{\dot{M}}{4\pi H R u^r m_{\rm p}} \simeq
\dfrac{c^2}{GM\sigma_T} \dfrac{R}{H} \dfrac{\mdot}{u^r}
\end{equation}
that implies the effective optical depth of
\begin{equation}
\begin{array}{l}
\displaystyle  \tau_{\rm eff} \simeq \tau_z
\sqrt{\frac{\sigma_{ff}}{\sigma_{\rm T}}} \simeq \\
\qquad{} \displaystyle  \simeq 5\times 10^{-3} \dfrac{\mdot^{3/2} E}{r^2
(u^r)^{3/2}} \left(\frac{H}{R}\right)^{-1/2}
\left(\frac{T}{\keV}\right)^{-1/4} \left(\frac{E}{\keV}\right)^{-3/2}.
\end{array}
\end{equation}
For a broad range of parameter values ($E\sim T
\sim T_{\rm max}$) including the critical mass accretion
rates estimated in this work, the effective optical depth is $\tau_{\rm eff}
\ll 1$. 

}

\section{Discussion and conclusions}\label{sec:disc}

\subsection{Limits of the standard disc approximation}\label{sec:limits}

We considered the balance of forces acting upon a test particle on the surface
of a thick disc. Apart from radiation pressure due to electron scattering
there are other forces that can induce launching winds from an accretion disc
including magnetic stresses and radiation pressure in emission lines
(see \citet{progawinds} for review about line-driven disc winds in active
galactic nuclei) hence it is probable that
in fact any disc is forming outflows even below the Eddington limit. However,
the continuum (at least, electron-scattering) optical depth of the outflow 
may become considerably high only above the critical accretion rate limit
(relevant optical depth estimates are given by~\citet{poutanen07}, equations
(29) and (31)). 

One of the main questions in the studies of supercritical accretion is the
relative importance of energy advection and matter loss. The two limiting
cases are the outflow-regulated accretion model introduced by \citet{SS73} and
the totally conservative slim disc model (see \citet{slim} for review) in
which the larger part of thermal energy released by viscous stresses is
advected inwards and finally lost in the black hole. Both models predict lower
accretion efficiency (per unit mass in-flowing at large distance) and
luminosity exceeding the Eddington limit by a logarithmic factor. Similar
logarithmic factor $\ln \left(\dfrac{R_{\rm out}}{R_{\rm in}}\right)$ arises
if one integrates the locally Eddington flux $\propto r^{-2}$ upon the surface
of a disc with the outer and inner radii equal to $R_{\rm out}$ and $R_{\rm
  in}$. Another effect usually neglected in accretion disc calculations but
probably important for thick discs is radial radiation diffusion. This process
tends to redistribute the radiation flux according to von~Zeipel rule
\citep{vonzeipel} $F\propto g$ (flux proportional to local gravity) that
scales identically with the local Eddington limit. Hence there are
different reasons for supercritical discs to attain the $T_{\rm eff} \propto
r^{-1/2}$ asymptotic for the super-critical regime and the following
logarithmic luminosity growth and ``flat'' spectrum ($\nu L_\nu \simeq
const$), and individual effects are difficult to disentangle. 

We find that transition to supercritical regime already appears for
considerably large advection: only about 10 per cent of the energy released in
the disc can escape as radiation and involved in acceleration of disc
winds. This justifies the use of models like \citet{lipunova99} and \citet{poutanen07} that
include both outflows and advection. In such models, the mass accretion rate
upon the black hole is considerably lower but still of the same order with the
inflow rate at the outer rim of the disc. The overall radiative efficiency of
mass accretion is probably close to 0.01$\div$0.02 with respect to the mass
absorbed by the black hole.

{

\subsection{Comparison to the slim disc model }\label{sec:disc:slim}

Our consideration still 
does not account for some other effects that become important
close to the limit. Firstly, the transonic nature of the thick disc solution
that is incorporated in the slim disc solution was not considered here. 
On the other hand, the slim-disc model does not account for the
vertical structure but rather operates with vertically-averaged equations. In
\citet{svert}, the vertical structure of a slim-disc was considered. 
However, the vertical structure itself
was in this paper assumed to be decoupled
from the radial structure of the disc, the vertical epicyclic frequency
(corresponding to $\Omega_z$ or $\Omega_\mu$ of the present work) was assumed
to be
constant with the vertical coordinate $z$, as well as all the velocity
components, that means that the disc was in fact treated as thin. This allows
to view our model as complementary to the slim disc. 

Another difference with the slim-disc approach is our unconventional 
treatment of advection effects. If advection is the primary term that breaks
the local equilibrium between dissipated and radiated energy, and the
advection effect is small, it is natural to describe advection as a delay term
in the energy conservation equation:
\begin{equation}
Q^+ - Q^- = (\vector{v} \nabla) U,
\end{equation}
where $U$ is the local surface energy density, and $Q^+$ and $Q^-$ are the
flux generated inside the disc and lost by its surface. One can approximate
$Q^- \simeq U/t_{\rm diff}$ and arrive to the approximate relation 
\begin{equation}
Q^+ \simeq \left( 1 + \dfrac{v_r}{t_{\rm diff}} \pardir{r}{}\right) Q^-,
\end{equation}
or, in a linear approximation,
\begin{equation}
Q^+(r+\delta r) \simeq Q^-(r),
\end{equation}
where $\delta r$ is the radial correction introduced in
section~\ref{sec:thick:adv}.
Besides, this approach has the clear physical meaning that the photons
generated at some radius drift downstream before being radiated. Of course,
this is a linear approach. For strong advection, the classical treatment of
advection as storage of some fixed part of the locally generated heat may
prove to be more relevant. In this sense, we also find our model complementary to
the slim-disc model. 

}

\subsection{Observational implications}\label{sec:imp}

It is often adopted for simplicity that the properties of accretion discs
change abruptly above the critical mass accretion rate. However, it seems that
for a large span of dimensionless mass accretion rates, approximately between
1 and 100, accretion may be considered close to standard but already altered
by thickness and advection. Besides, for $\mdot\gtrsim 1$, the optically thick
part of the disc extends further inwards that is extremely important in
continuum fitting. Detailed continuum fitting is important for estimating
black hole spins \citep{lifits,NMC12}. Black hole spins are measured in the
high, or soft, states associated with higher mass accretion rates with
$\mdot \sim 1$ \citep{MM00}. For $\mdot=1$, corrections for vertical gravity
and advection are unimportant, but rapidly increase with \mdot. In future,
more comprehensive models they should be taken into account together with
the effects of radiation transfer and scattering in the inner parts of
the flow. Since the inner disc edge is effectively shifted closer to the event
horizon one should expect ignoring the radial advection to overestimate the
spin of the black hole.

Bright quasars and QSO (quasi-stellar objects) could be another group of
objects that exist close to the criticality limit. The puzzling five per cent
accretion efficiency required by some observational data such as quasar
microlensing \citep{morgan10} may be
connected to mild advection effects lowering the radiative efficiency of
QSO accretion discs. 

\subsection{Conclusions}\label{sec:conc}

We summarize that, due to stronger vertical gravity, 
transition to super-Eddington accretion should appear at
higher mass accretion rates. 
Close to the local Eddington limit, disc thickness becomes large and several
effects neglected in the standard disc model become important. Here, we
estimate the effects of vertical gravity dependence on vertical coordinate and
the effects of non-local emission of the energy released in the disc. The
latter is primarily connected to radial advection. We find that including
both effects simultaneously tends to further increase the local Eddington
limit, but the efficiency of accretion decreases to about one-two per cent for
the critical mass accretion rate. More comprehensive
studies taking into account the non-one-dimensional character of thick
accretion disc structure are needed to decide upon the relative importance of
different effects such as advection and outflow formation in accretion discs
exceeding the Eddington limit. It may turn in fact that there is no unique
Eddington limit, but accretion disc properties change much more gradually with
mass accretion rate. 


\section*{Acknowledgments}

We wish to thank Juri Poutanen and
Nikolai Shakura for valuable discussions. PA thanks Dynasty Foundation and
Academy of Finland grant 268740 for
support, AC thanks Finnish Centre for
MObility Exchange (CIMO). We also acknowledge support from Russian Foundation
for Basic Research grant 14-02-91172. 

\bibliographystyle{mn2e}
\bibliography{mybib}

\appendix

\section{Keplerian von Zeipel cylinders in Kerr metric}\label{sec:app:KvZ}

\subsection{Basic equations}

Let us assume that dimensionless 
angular momentum $l=-u_\varphi / u_t$ and angular velocity
$\omega = u^\varphi / u^t$ depend on each other unambiguously,
$l=l(\omega)$. Note that $l$ here is the angular momentum {\it per unit
  energy} rather than per unit rest mass. 
This is the case for barotropic flows where gradients of
pressure and density are collinear
everywhere. This sentence is known as Poincar\'e-Wavre theorem and is proven both for
classical mechanics (see~\citet{Tassoul}) and for General Relativity
\citep{FS13}. Sometimes it is also referred to as von Zeipel theorem
\citep{relvonzeipel, vzc}. Surfaces of constant $l$ and $\omega$ coincide \citep{vzc}
and are known as von Zeipel cylinders. Their shapes
however rely upon the given relation between $l$ and $\omega$. 

Let us assume that the accretion disc is strictly Keplerian in the equatorial
plane:
\begin{equation}\label{E:omegaK}
\omega = \omega_K := \dfrac{1}{r^{3/2}+a},
\end{equation}
\begin{equation}\label{E:lK}
l = l_K := \dfrac{r^2-2a\sqrt{r}+a^2}{r^{3/2}-2\sqrt{r}+a}.
\end{equation}
Expression (\ref{E:omegaK}) allows to exclude the radius
$r=\left(\omega^{-1}-a\right)^{2/3}$ and evaluate angular momentum as a
function of $\omega$: 
\begin{equation}\label{E:lo}
l(\omega) =
\dfrac{\left(1-a\omega\right)^{1/3}\left(1-3a\omega\right)+a^2\omega^{4/3}}{\omega^{1/3}\left(1-2\left(1-a\omega\right)^{1/3}\omega^{2/3}\right)}.
\end{equation}
Angular velocity may be then found as a function of the two spatial
coordinates by solving the equation:
\begin{equation}
l(\omega) = - \dfrac{g_{\varphi t} + \omega g_{\varphi \varphi}}{g_{tt}+\omega
g_{\varphi t}},
\end{equation}
where the left-hand side applies (\ref{E:lo}), while the right-hand side is a
consequence of velocity conversion relations $u_\varphi = g_{\varphi t} u^t +
g_{\varphi \varphi} u^\varphi$, $u_t = g_{t t} u^t + g_{\varphi t}
u^\varphi$. Relevant metric coefficients:
\begin{equation}
g_{tt} = -1 + \dfrac{2r}{\rho^2},
\end{equation}
\begin{equation}
g_{\varphi t} = -\dfrac{2ar(1-\mu^2)}{\rho^2},
\end{equation}
\begin{equation}
g_{\varphi \varphi} =  \dfrac{\Sigma^2}{\rho^2} \left(1-\mu^2\right),
\end{equation}
where:
\begin{equation}
\mu^2 = \cos^2\theta,
\end{equation}
\begin{equation}
\rho^2 = r^2+a^2\mu^2,
\end{equation}
\begin{equation}
\Sigma^2 = (r^2+a^2)^2 - a^2 \Delta \left(1-\mu^2\right),
\end{equation}
\begin{equation}
\Delta = r^2 + a^2 - 2 r.
\end{equation}

Finally, the main equation for angular velocity becomes
\begin{equation}\label{E:leq}
\dfrac{\left(1-a\omega\right)^{1/3}\left(1-3a\omega\right)+a^2\omega^{4/3}}{\omega^{1/3}\left(1-2\left(1-a\omega\right)^{1/3}\omega^{2/3}\right)}
=
\dfrac{\Sigma^2\omega - 2ar}{\dfrac{\rho^2-2r}{1-\mu^2}+2ar\omega}.
\end{equation}

\subsection{Schwarzschild case}\label{E:Sch}

\begin{figure}
 \centering
\includegraphics[width=\columnwidth]{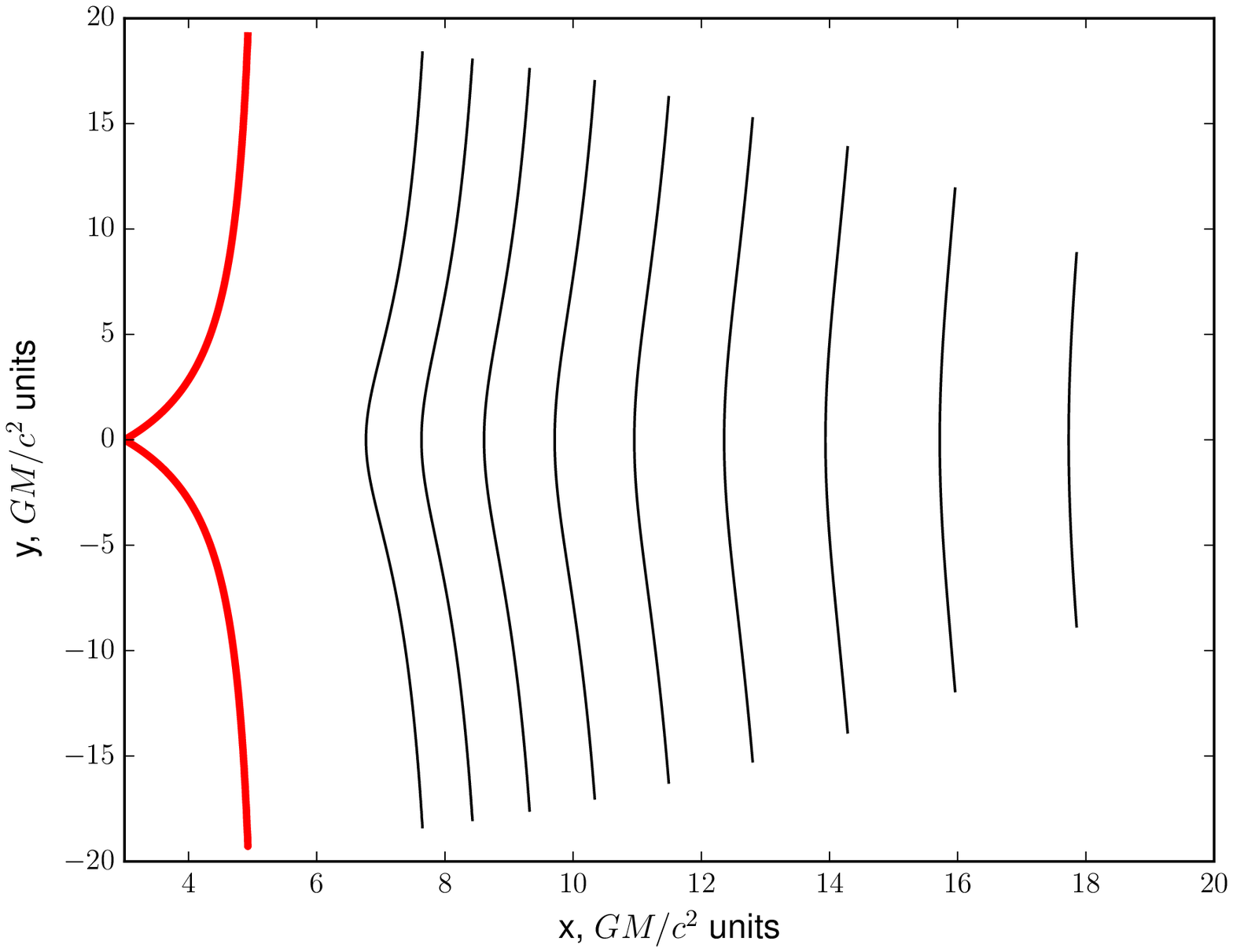}
\caption{Keplerian von Zeipel cylinders for a Schwarzschild black hole. Red
  thick line is the applicability limit of the solution given by~(\ref{E:cube:sch}).
}
\label{fig:app:Sch}
\end{figure}

If one sets $a=0$, it is possible to obtain an analytical expression for
angular velocity as a function of $r$ and $\mu$. Equation (\ref{E:leq})
becomes
\begin{equation}
\omega^{4/3}  \left( 1-2\omega^{2/3}\right) = \dfrac{r-2}{(1-\mu^2)r^3}.
\end{equation}
This equation is cubic both in $r$ (useful for recovering the shapes of the
cylinders) and in $f=\omega^{2/3}$.
\begin{equation}
f^3-\dfrac{1}{2}f^2 + X = 0,
\end{equation}
where $X=\dfrac{r-2}{2(1-\mu^2)r^3}$. This equation is easily transformed into
a depressed cubic equation $t^3+pt+q=0$ by a substitute $t=f-1/6$:
\begin{equation}
t^3-\dfrac{1}{12}t + \left( X - \dfrac{1}{108}\right)=0.
\end{equation}
In this equation, $p=-1/12<0$. Equation discriminant $D =
4p^3 + 27 q^2 <0$ for any $r>6$ and $\mu < 1/\sqrt{2}$, hence:
\begin{equation}\label{E:cube:sch}
t=\dfrac{1}{3} \cos\left( \dfrac{1}{3} \acos\left( 1-108X \right)-\dfrac{2\pi}{3}\right),
\end{equation}
and $\omega = \left(t+\dfrac{1}{6} \right)^{3/2}$. Keplerian rotation field in
the Schwarzschild case is shown in Fig.~(\ref{fig:app:Sch}). 

\subsection{General Kerr case: solution and results}

\begin{figure*}
 \centering
\includegraphics[width=\textwidth]{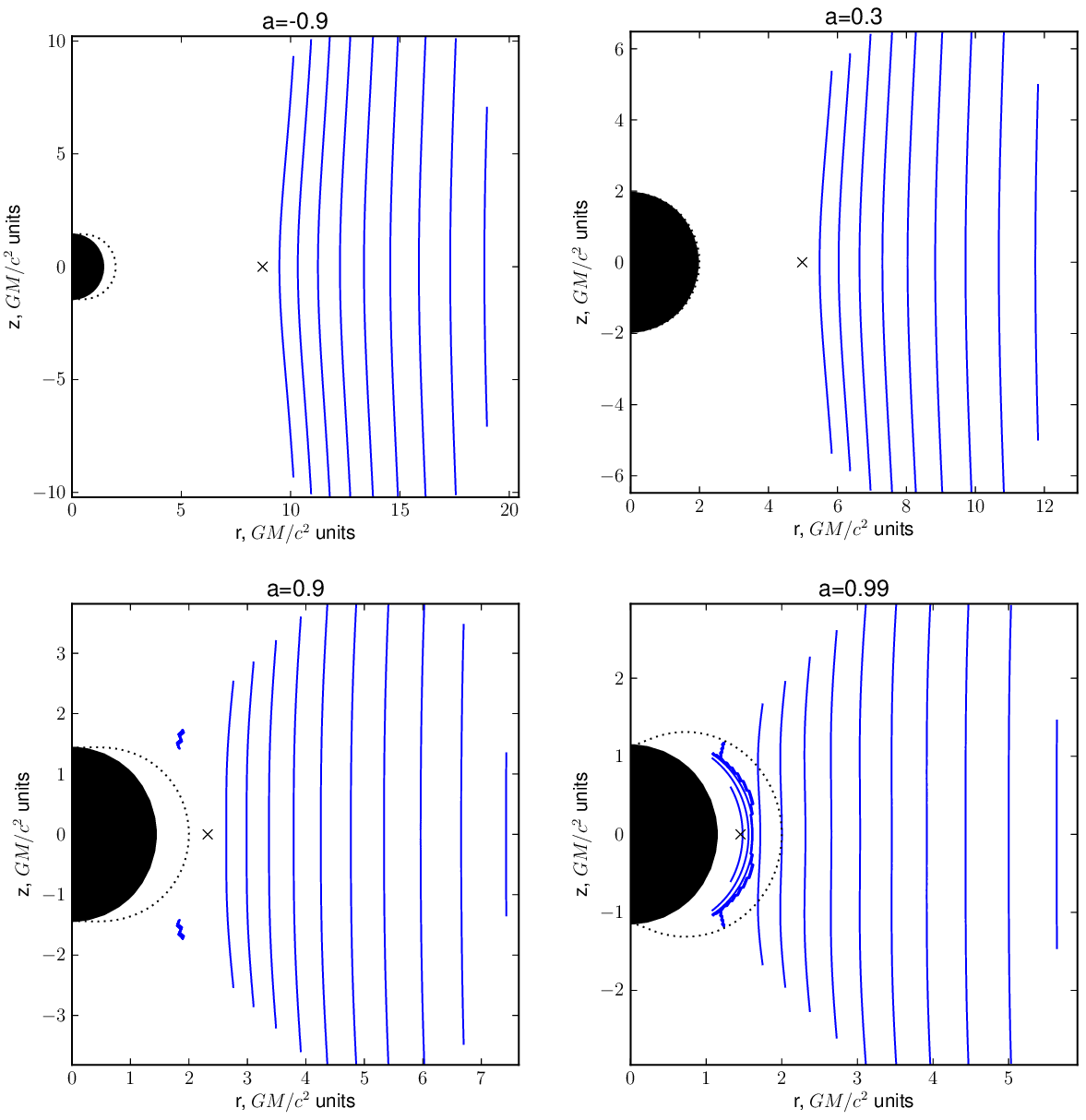}
\caption{Keplerian von Zeipel cylinders for different Kerr parameters. Dotted
  curves mark the outer limit of the ergosphere. Cross is the position of the
  last stable orbit in the equatorial plane. 
}
\label{fig:app:os}
\end{figure*}

Right-hand side of equation~(\ref{E:leq}) is singular at small $\omega$, 
but when searching for the solution for given $r$ and $\mu$, one
should be conscious about the singularity in the left-hand side of the
equation, setting the maximal possible frequency:
\begin{equation}
1-2\left(1-a\omega_{\rm max}\right)^{1/3}\omega_{\rm max}^{2/3}=0,
\end{equation}
\begin{equation}
\omega_{\rm max}^3 - \dfrac{1}{a} \omega_{\rm max}^2 + \dfrac{1}{8a}=0
\end{equation}
Substituting $\omega_{\rm max}=t+\dfrac{1}{3a}$, one arrives again to a cubic
equation:
\begin{equation}
t^3 - \dfrac{1}{3a^2} t - \dfrac{2}{27 a^3} + \dfrac{1}{8a} =0.
\end{equation}
For any $a\neq 0$, $p=-\dfrac{1}{3a^2} < 0$, $D = \dfrac{4}{27 a^6}\left(
1+\left(1-\dfrac{27}{16}a^2\right)^2\right)>0$ hence the solution may be
expressed as
\begin{equation}
\begin{array}{l}
\displaystyle \omega_{max} = - \frac{2}{3a} \sign\left( \dfrac{1}{8a}-\dfrac{2}{27
  a^3}\right)\times\\
\qquad{}\displaystyle \times
\cosh\left(\dfrac{1}{3}\acosh\left(\dfrac{3}{16}\left|a\left(1-\dfrac{16}{27a^2}\right)\right|\right)
\right)\\
\end{array}
\end{equation}
For $0<\omega<\omega_{\rm max}$, equation~(\ref{E:leq}) was solved numerically for
$\omega$ for different $r$ outside the last stable orbit radius and for
$|\mu|< \dfrac{1}{\sqrt{2}}$. Isolines of resulting angular frequency for
different Kerr parameter values are shown in Fig.~\ref{fig:app:os}.

\label{lastpage}

\end{document}